\PassOptionsToPackage{unicode}{hyperref}
\PassOptionsToPackage{hyphens}{url}
\PassOptionsToPackage{dvipsnames,svgnames,x11names}{xcolor}
\documentclass[
  12pt]{article}

\usepackage{amsmath,amssymb,amsthm}
\usepackage{iftex}
\ifPDFTeX
  \usepackage[T1]{fontenc}
  \usepackage[utf8]{inputenc}
  \usepackage{textcomp} 
\else 
  \usepackage{unicode-math}
  \defaultfontfeatures{Scale=MatchLowercase}
  \defaultfontfeatures[\rmfamily]{Ligatures=TeX,Scale=1}
\fi
\usepackage{lmodern}
\ifPDFTeX\else  
\fi
\IfFileExists{upquote.sty}{\usepackage{upquote}}{}
\IfFileExists{microtype.sty}{
  \usepackage[]{microtype}
  \UseMicrotypeSet[protrusion]{basicmath} 
}{}
\makeatletter
\@ifundefined{KOMAClassName}{
  \IfFileExists{parskip.sty}{%
    \usepackage{parskip}
  }{
    \setlength{\parindent}{0pt}
    \setlength{\parskip}{6pt plus 2pt minus 1pt}}
}{
  \KOMAoptions{parskip=half}}
\makeatother
\usepackage{xcolor}
\makeatletter
\ifx\paragraph\undefined\else
  \let\oldparagraph\paragraph
  \renewcommand{\paragraph}{
    \@ifstar
      \xxxParagraphStar
      \xxxParagraphNoStar
  }
  \newcommand{\xxxParagraphStar}[1]{\oldparagraph*{#1}\mbox{}}
  \newcommand{\xxxParagraphNoStar}[1]{\oldparagraph{#1}\mbox{}}
\fi
\ifx\subparagraph\undefined\else
  \let\oldsubparagraph\subparagraph
  \renewcommand{\subparagraph}{
    \@ifstar
      \xxxSubParagraphStar
      \xxxSubParagraphNoStar
  }
  \newcommand{\xxxSubParagraphStar}[1]{\oldsubparagraph*{#1}\mbox{}}
  \newcommand{\xxxSubParagraphNoStar}[1]{\oldsubparagraph{#1}\mbox{}}
\fi
\makeatother

\usepackage{longtable,booktabs,array}
\usepackage{calc} 
\usepackage{etoolbox}
\makeatletter
\patchcmd\longtable{\par}{\if@noskipsec\mbox{}\fi\par}{}{}
\makeatother
\IfFileExists{footnotehyper.sty}{\usepackage{footnotehyper}}{\usepackage{footnote}}
\makesavenoteenv{longtable}
\usepackage{graphicx}
\makeatletter
\def\maxwidth{\ifdim\Gin@nat@width>\linewidth\linewidth\else\Gin@nat@width\fi}
\def\maxheight{\ifdim\Gin@nat@height>\textheight\textheight\else\Gin@nat@height\fi}
\makeatother
\setkeys{Gin}{width=\maxwidth,height=\maxheight,keepaspectratio}
\makeatletter
\def\fps@figure{htbp}
\makeatother

\makeatletter
\@ifpackageloaded{caption}{}{\usepackage{caption}}
\AtBeginDocument{%
\ifdefined\contentsname
  \renewcommand*\contentsname{Table of contents}
\else
  \newcommand\contentsname{Table of contents}
\fi
\ifdefined\listfigurename
  \renewcommand*\listfigurename{List of Figures}
\else
  \newcommand\listfigurename{List of Figures}
\fi
\ifdefined\listtablename
  \renewcommand*\listtablename{List of Tables}
\else
  \newcommand\listtablename{List of Tables}
\fi
\ifdefined\figurename
  \renewcommand*\figurename{Figure}
\else
  \newcommand\figurename{Figure}
\fi
\ifdefined\tablename
  \renewcommand*\tablename{Table}
\else
  \newcommand\tablename{Table}
\fi
}
\@ifpackageloaded{float}{}{\usepackage{float}}
\floatstyle{ruled}
\@ifundefined{c@chapter}{\newfloat{codelisting}{h}{lop}}{\newfloat{codelisting}{h}{lop}[chapter]}
\floatname{codelisting}{Listing}

\makeatother
\makeatletter
\@ifpackageloaded{caption}{}{\usepackage{caption}}
\@ifpackageloaded{subcaption}{}{\usepackage{subcaption}}
\makeatother

\ifLuaTeX
  \usepackage{selnolig}  
\fi
\usepackage[]{natbib}
\usepackage{bookmark}

\IfFileExists{xurl.sty}{\usepackage{xurl}}{} 
\hypersetup{
  pdftitle={Title},
  pdfauthor={Author 1; Author 2},
  pdfkeywords={3 to 6 keywords, that do not appear in the title},
  colorlinks=true,
  linkcolor={blue},
  filecolor={Maroon},
  citecolor={Blue},
  urlcolor={Blue},
  pdfcreator={LaTeX via pandoc}}

\theoremstyle{definition}
\newtheorem{defn}{\protect\definitionname}
\theoremstyle{definition}
\newtheorem{example}{\protect\examplename}

\usepackage{accents}

\makeatother

\usepackage{babel}
\providecommand{\definitionname}{Definition}
\providecommand{\examplename}{Example}

\newcommand{\anon}{1}

\begin{document}

\def\spacingset#1{\renewcommand{\baselinestretch}%
{#1}\small\normalsize} \spacingset{1}


\if1\anon
{
  \title{\bf Semi-tail Units: A Universal Scale for Test Statistics and Efficiency}
  \author{Paul W. Vos\hspace{.2cm}\\
    Department of Public Health, East Carolina University}
  \maketitle
} \fi

\if0\anon
{
  \bigskip
  \bigskip
  \bigskip
  \begin{center}
    {\LARGE\bf Title}
\end{center}
  \medskip
} \fi

\bigskip
\begin{abstract}
We introduce $\zeta$- and $s$-values as quantile-based standardizations that are
particularly suited for hypothesis testing. Unlike p-values, which
express tail probabilities, $s$-values measure the number of semi-tail
units into a distribution's tail, where each unit represents a halving
of the tail area. This logarithmic scale provides intuitive interpretation:
$s=3.3$ corresponds to the 10th percentile, $s=4.3$ to the 5th percentile,
and $s=5.3$ to the 2.5th percentile. For two-tailed tests, $\zeta$-values
extend this concept symmetrically around the median.

We demonstrate how these measures unify the interpretation of all
test statistics on a common scale, eliminating the need for distribution-specific
tables. The approach offers practical advantages: critical values
follow simple arithmetic progressions, combining evidence from independent
studies reduces to the addition of $s$-values, and semi-tail units provide
the natural scale for expressing Bahadur slopes. This leads to a new
asymptotic efficiency measure based on differences rather than ratios
of slopes, where a difference of 0.15 semi-tail units means that the more
efficient test moves samples 10\% farther into the tail. Through examples
ranging from standardized test scores to poker hand rankings, we show
how semi-tail units provide a natural and interpretable scale for
quantifying extremeness in any ordered distribution. 

\end{abstract}

\noindent%

\vfill

\newpage

\spacingset{1.8} 

\section{Standardization}

\label{i.-standardization}

\subsection{z-score}

\label{z-score}

The $z$-score, also known as the standard score, is perhaps the most
widely used form of standardization in statistics. For a distribution
$X$ with mean $\mu$ and standard deviation $\sigma$, the $z$-score
for a value $x\in X$ is: $z=\frac{x-\mu}{\sigma}$. This transformation
produces a standardized distribution with mean 0 and standard deviation
1, provided the first two moments exist. The primary utility of $z$-scores
lies in their ability to place values from different distributions
on a common scale, enabling meaningful comparisons, particularly when
the distributions belong to the same location-scale family.

Consider comparing performance on two standardized tests: the SAT
and ACT. The SAT has a mean score of approximately 1500 with a standard
deviation of 300, while the ACT has a mean of 21 with a standard deviation
of 5. Without standardization, comparing a score of 1720 on the SAT
with a score of 27 on the ACT would be difficult. Converting to $z$-scores
reveals that an SAT score of 1720 yields $z=\frac{1720-1500}{300}=0.73$,
while an ACT score of 27 yields $z=\frac{27-21}{5}=1.2$.

Thus, the SAT score is 0.73 standard deviations above the mean, while
the ACT score is 1.2 standard deviations above the mean. If both distributions
have the same shape, the ACT score of 27 lies farther in the tail
than the SAT score of 1720. When both distributions are normal, these
$z$-values convert directly to percentiles: a $z$-score of 0.73
corresponds to approximately the 77th percentile, while a $z$-score
of 1.2 corresponds to approximately the 88th percentile.

\subsection{Affine Transformations}

The $z$-score exemplifies an important invariance principle: meaningful
statistical methods must be invariant under affine transformations
of the data. An affine transformation takes the form $Y=aX+b$ where
$a\neq0$. The collection of all such transformations forms a group
under composition that preserves essential distributional properties.

Standardized moments are invariant under affine transformations. The
$k$th standardized moment of $X$ is 
\[
\tilde{\mu}_{k}(X)=\sigma^{-k}E(X-\mu)^{k}
\]
where $\mu=EX$ and $\sigma^{2}=E(X-\mu)^{2}$. For all $a\neq0$,
$\tilde{\mu}_{k}(aX+b)=\tilde{\mu}_{k}(X)$.

This invariance ensures that statistical conclusions do not depend
on arbitrary measurement units. Whether we measure temperature in
Celsius or Fahrenheit, distance in meters or feet, or currency in
dollars or euros, the standardized properties remain unchanged.

\begin{figure}
\centering{
\includegraphics{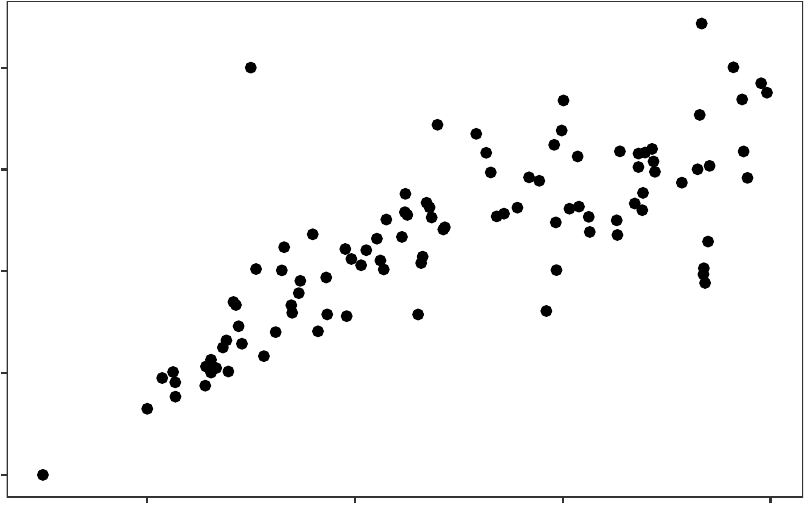}}
\caption{Scatter plot showing preservation of features under affine transformation.
Axes labels are not shown as these are the only parts of the graph
that would change under affine transformations.\label{fig:scatter}}
\end{figure}

The scatter plot in Figure \ref{fig:scatter} illustrates how important data features beyond
moments are retained under affine transformations. The plot displays
an outlier, a high leverage point, slight nonlinearity, and increasing
variance. For every affine transformation of the variables, these features
remain identifiable---only the axis labels change. 

\subsection{Quantile Standardization}

\label{quantile-standardization}

While $z$-score standardization preserves distribution shape, quantile
standardization takes a different approach by mapping quantiles of
one distribution to those of a reference distribution.

Given a distribution $X$ with cumulative distribution function (CDF)
$F_{X}$ and a reference distribution with CDF $F_{R}$, quantile
standardization transforms a value $x\in X$ to: 
\[
y=F_{R}^{-1}\circ F_{X}(x)
\]

When both $X$ and $R$ are continuous, this transformation ensures
that the resulting distribution has CDF $F_{R}$. When $X$ is discrete,
the transformed distribution $Y$ need not exactly match the reference
distribution. Nevertheless, the values $y=F_{R}^{-1}(F_{X}(x))$ provide
useful information about the position of $x$ in the lower tail of
distribution $X$.

\subsection{Strictly Monotone Transformations}

\label{quantile-standardization-and-strictly-monotone-transformations}

Quantile standardization of a continuous distribution exemplifies
a monotone transformation. A transformation $g$ is monotone if $g(x_{1})<g(x_{2})$
whenever $x_{1}<x_{2}$ (strictly increasing) or $g(x_{1})>g(x_{2})$
whenever $x_{1}<x_{2}$ (strictly decreasing). The set of monotone
transformations on $\mathbb{R}$ forms a group under composition.

Properties of $X$ that are invariant under affine transformations
are not preserved under monotone transformations. (The affine transformations
form a proper subgroup of the monotone transformations.) When distributional
shape is important, monotone transformations may lose crucial information.
What remains invariant is the ordering of observations (reversed for
decreasing transformations), making monotone transformations well-suited
when order is the primary concern.

The distinction between these standardization approaches---$z$-scores
preserving shape via affine transformation versus quantile standardization
preserving order via monotone transformation---motivates the development
of $\zeta$- and $s$-scores as standardizations particularly suited
for test statistics.

\section[zeta and s scores]{\(\zeta\)- and \(s\)-scores}

\subsection{Definitions}

\label{definitions}

The $\zeta$-score provides a quantile-based standardization particularly
well-suited for test statistics and applications where tail behavior
is of primary interest. As is true for $z$-scores, the terms $\zeta$-score
and $\zeta$-value indicate the same quantity. 
\begin{defn}
For distribution $T$, the\emph{ $\zeta$-value of} $t\in T$ is 
\[
\zeta=\zeta(t)=\text{sign}(T_{L}(t)-T_{R}(t))\log_{2}\min\{2T_{L}(t),2T_{R}(t)\}
\]
where $T_{L}$ and $T_{R}$ are the left and right tails: $T_{R}(t)=\min\{1/2,P(T\geq t)\}$
and $T_{L}(t)=\min\{1/2,P(T\leq t)\}$. The sign of $\zeta$ indicates
the tail, and the tail size is $2^{-|\zeta|-1}$. When only one tail
is of interest, we use the $s$-\emph{value}, defined by 
\[
s=|\zeta|+1.
\]
\end{defn}
Moving one unit farther into the tail halves the tail proportion:
$P(S\ge s+1)=\frac{1}{2}P(S\ge s)$ for any value $s$. This property
also holds for $\zeta$-values where in the left tail: $P(Z\le\zeta-1)=\frac{1}{2}P(Z\le\zeta)$.
As the $z$-score measures standard deviation units from the mean,
the $\zeta$-score measures \emph{semi-tail units} from the median.
The term ``semi-tail unit'' refers to a unit increment on the logarithmic
scale, not a unit of measurement in the dimensional sense. Just as
the Richter scale measures earthquake magnitude in dimensionless units
where each unit increase represents a tenfold increase in amplitude,
each semi-tail unit represents a halving of the tail proportion. 

Let $S$ denote the distribution of $s$-values and $Z$ the distribution
of $\zeta$-values. The median of $S$ is 1 while the median of $Z$
is 0. When $T$ is continuous, $Z$ follows the Laplace distribution
with MAD (median absolute deviation) 1 and $S$ follows the exponential
distribution, both with scale parameter $(\log2)^{-1}$.

\subsection{Reference Values}

A single reference value, $s=3.3$, along with the halving property
provide several useful benchmarks:
\begin{itemize}
\item $s=3.3$ corresponds to the upper 10\% ($2^{-3.3}\approx0.10$) 
\item Adding 1 gives $s=4.3$: upper 5\% ($2^{-4.3}\approx0.05$) 
\item Adding $3.3$ gives $s=6.6$: upper 1\% ($0.10^{2}=0.01$)
\item Subtracting 1 gives $s=5.6$: upper 2\% ($2^{-5.6}\approx0.02$) 
\end{itemize}
Continuing to add and subtract unit semi-tail values provides other
useful benchmarks. For example, $s=6.6,5.6,4.6,$ and $3.6$ are the
upper 1, 2, 4, and 8 percent tails, respectively. 

\subsection{Examples}

We illustrate $\zeta$-scores by extending the SAT/ACT example from
Section 2. 
\begin{example}
\textbf{Test Scores.} If SAT and ACT scores follow normal distributions
with the parameters given earlier, an SAT score of 1720 has $\zeta=1.1$
while an ACT score of 27 has $\zeta=2.1$. This places 1720 at 1.1
semi-tail units into the right tail of SAT scores, while 27 is 2.1
semi-tail units into the right tail of ACT scores. Since $\zeta=1$
marks the third quartile, 1720 is in the top 25\% of SAT scores. Being
one semi-tail unit farther, 27 is in the top 12.5\% of ACT scores.

Now consider when the distributions have different shapes: SAT scores
remain normal but ACT scores follow a Cauchy distribution. Since the
Cauchy distribution lacks finite moments, we use the median and MAD
for location and scale. For a normal distribution, the median equals
the mean and MAD $=Q_{3}\sigma$ where $Q_{3}=0.6745$ is the third
quartile of the standard normal.

Using these measures, the $z$-scores become $0.49=0.6745\times0.73$
and $0.81=0.6745\times1.2$ for the SAT and ACT scores respectively.
The $\zeta$-score for the SAT score remains $\zeta=1.1$ (top 24\%).
The $\zeta$-score for the ACT score is now $\zeta=1.6$. Using the
reference $s=3.6$ (upper 8\%), we find that 27 is in the top 16\%
of ACT scores. 
\end{example}
The next example illustrates a distribution where only the upper tail
is relevant. 

\begin{example}
\textbf{5-card Poker Hands.} Poker uses an ordering of hands from
a standard 52-card deck. Cards are ranked from highest to lowest:
Ace (A), King (K), Queen (Q), Jack (J), Ten (T), 9, 8, ..., 2. Five-card
hands are ordered by repeated values (pairs, triples, full house,
four of a kind), suits (flush), and consecutive order (straight, straight
flush).

Table \ref{tab:Distribution-of-ranks} shows the distribution of hand
rankings. The highest hand, a royal flush (AKQJT of one suit), occurs
4 times among all possible hands. The Count column lists the number
of each hand type, Tail Count shows cumulative counts, Tail Prop shows
cumulative proportions, and the final column shows semi-tail units.
Note that $s$-values from 1 to 19.3 span five orders of magnitude
in probability.

The semi-tail values reveal useful relationships. The median hands
are AKQJ7 and AKQJ6 (non-flush) with ranks 6188 and 6189. Since $2^{-3.3}\approx0.10$,
the lowest two pair ($s=3.7$) is beaten by 10 times as many hands
as the lowest straight ($s=7.0=3.7+3.3$). The lowest straight is
beaten by more than twice as many hands as the lowest flush ($s=8.1>7.0+1$),
which in turn is beaten by twice as many hands as the lowest full
house ($s=9.2>8.1+1$).

If both tails were relevant, we would use $\zeta$-values. The median
hands have $\zeta=0$. Upper tail ranks have $\zeta=s-1$. For lower
tail ranks, the lowest hand (75432 non-flush, occurring $4^{5}-4=1020$
times) has $\zeta=\log_{2}(1020/\genfrac(){0pt}{2}{52}{5})=-11.3$. The 2nd,
4th, and 8th lowest hands have $\zeta=-10.3,-9.3,$ and $-8.3$ respectively,
demonstrating the halving property in the lower tail. 
\end{example}
\begin{table}
\begin{tabular}{|c|c|r|r|r|r|}
\hline 
Hand & Ranks & Count & Tail Count & Tail Prop. & s\tabularnewline
\hline 
\hline 
Royal Flush & 1 & 4 & 4 & 0.00000154 & 19.3\tabularnewline
\hline 
Straight Flush & 2 to 10 & 36 & 40 & 0.00001539 & 16.0\tabularnewline
\hline 
Four of a Kind & 11 to 166 & 624 & 664 & 0.00025549 & 11.9\tabularnewline
\hline 
Full House & 167 to 322 & 3,744 & 4,408 & 0.00169606 & 9.2\tabularnewline
\hline 
Flush & 323 to 1599 & 5,108 & 9,516 & 0.00366146 & 8.1\tabularnewline
\hline 
Straight & 1600 to 1609 & 10,200 & 19,716 & 0.00758611 & 7.0\tabularnewline
\hline 
Three of a Kind & 1610 to 2467 & 54,912 & 74,628 & 0.02871456 & 5.1\tabularnewline
\hline 
Two pair & 2468 to 3325 & 123,552 & 198,180 & 0.07625358 & 3.7\tabularnewline
\hline 
Pair & 3326 to 6185 & 1,098,240 & 1,296,420 & 0.49882261 & 1.0\tabularnewline
\hline 
AKQJ9 not flush & 6186 & 1,020 & 1,297,440 & 0.49921507 & 1.0\tabularnewline
\hline 
AKQJ8 not flush & 6187 & 1,020 & 1,298,460 & 0.49960754 & 1.0\tabularnewline
\hline 
AKQJ7 not flush & 6188 & 1,020 & 1,299,480 & 0.50000000 & 1.0\tabularnewline
\hline 
Other High Card & 6189 to 7462 & 1,299,480 & 2,598,960 & 1.00000000 & 0.0\tabularnewline
\hline 
\end{tabular} \caption{\label{tab:Distribution-of-ranks}Distribution of ranks of 5-card
poker hands.}
\end{table}

\section{Semi-tail Tests}

\subsection{Test Statistics}

Test statistics are central to statistical inference. A test statistic
relates a sample to a claim about the population distribution. This
claim, called the null hypothesis, specifies the null distribution.
The test statistic assigns a numerical value to each sample, which
is then compared to its sampling distribution---the distribution of
the test statistic over all possible samples from the null distribution.

Extreme values of the test statistic indicate samples that are unlikely
under the null hypothesis. However, the variety of sampling distributions
(e.g., normal, binomial, $t$, $F$, chi-squared with various shape parameters)
makes it difficult to identify extreme observations without consulting
tables or software. Standardizing test statistics in semi-tail units
provides a universal scale for identifying extreme observations.

\subsection[Standardized Tests]{Standardized Tests: the $s$- and $\zeta$-tests}

Test statistics fall into two categories: one-tailed and two-tailed.
One-tailed tests treat large positive values as extreme (e.g., $F$-test,
chi-squared test), while two-tailed tests treat large absolute values
as extreme (e.g., $t$-test, $z$-test).

For one-tailed tests, the $F$-distribution's shape varies with numerator
and denominator degrees of freedom, making it difficult to know what
constitutes an extreme value. We standardize using the semi-tail value:
\begin{equation}
s_{obs}=-\log_{2}P(T_{\nu_{1},\nu_{2}}\ge t_{obs})\label{eq:sobs}
\end{equation}
where $T_{\nu_{1},\nu_{2}}$ follows the $F$-distribution with $\nu_{1}$
and $\nu_{2}$ degrees of freedom and $t_{obs}$ is the observed value
of the test $T_{\nu_{1},\nu_{2}}$ . All one-tailed tests can be similarly
standardized. When the sampling distribution is continuous, $S$ follows
the exponential distribution with median 1 and tail area: 
\[
P(S\ge s)=2^{-s}.
\]

Two-tailed tests use $\zeta$-values. For $t_{obs}>0$ from the $t$-distribution
with $\nu$ degrees of freedom: 
\[
\zeta_{obs}=-\log_{2}\left(2P(T_{\nu}\ge t_{obs})\right)=s_{obs}-1
\]
where $s_{obs}$ is defined as in (\ref{eq:sobs}). Let $Z$ denote
the distribution of $\zeta$-values under the null hypothesis. For
continuous two-tailed tests, $Z$ follows the Laplace distribution
with median zero and MAD 1, with double tail area: 
\begin{equation}
P(|Z|\ge|\zeta_{obs}|)=2^{-|\zeta_{obs}|}\label{eq:pZETA}
\end{equation}
since $|Z|$ has the same distribution as $S$.

For discrete distributions, equation (\ref{eq:pZETA}) may
not hold exactly. In practice, conditioning on the observed tail is
more meaningful: 
\begin{equation}
P\left(|Z|\ge|\zeta_{obs}|\Big|D=d_{obs}\right)=2^{-|\zeta_{obs}|}\label{eq:pZETA-1}
\end{equation}
where $d_{obs}=\pm1$ indicates the tail containing $\zeta_{obs}$.

Equation (\ref{eq:pZETA-1}) holds when $D(T)$ is uniform on $\{-1,1\}$; this is the case for $D$ defined as follows. Let $\mathcal{T}_{0}=\{t:P(T\le t)>1/2,P(T\ge t)>1/2\}$
be the set of values that could belong to either tail. For continuous
distributions, $\mathcal{T}_{0}$ is typically empty. When $P(\mathcal{T}_{0})=0$:
\[
D(t)=\begin{cases}
-1 & P(T\le t)\le\frac{1}{2}\\
\hphantom{-}1 & P(T\ge t)\le\frac{1}{2}\\
\hphantom{-}0 & \text{otherwise}
\end{cases}
\]

When $P(\mathcal{T}_{0})>0$, values in $\mathcal{T}_{0}$ are allocated
to ensure $D$ is uniform on $\{-1,1\}$: 
\[
D(t)=\begin{cases}
-1 & P(T\le t)\le\frac{1}{2}\\
\hphantom{-}1 & P(T\ge t)\le\frac{1}{2}\\
\hphantom{-}B & \text{otherwise}
\end{cases}
\]
where $B$ is Bernoulli with mass proportional to $\{1-2P(D=-1),1-2P(D=1)\}$.

When standardized to $\zeta$-values, nonempty $\mathcal{T}_{0}$
contains only $\zeta=0$, meaning samples in both tails are zero semi-tail
units into each tail. This apparent randomization is inconsequential:
whether we report the sample as zero units into the left or right
tail, the interpretation remains the same.

\subsection{Interpretation}

The interpretation of p-values has generated considerable confusion
\citep{Wasserstein29032019}. While some confusion stems from misleading
criticisms \citep{Greenland2019}, much arises from probability's
dual role in describing random outcomes and degrees of belief. The
p-value is conventionally interpreted using a hypothetical sequence of
samples from the null distribution, not the actual population, thereby adding
unnecessary complexity.

The $s$- and $\zeta$-values describe the relationship between a
sample and its sampling distribution using percentiles rather than
probabilities. While percentiles can be related to probabilities,
their definition requires no probability concepts. For example, a
six-foot-four-inch man is in the upper 2\% of US male heights\footnote{\url{https://dqydj.com/height-percentile-calculator-for-men-and-women/}}.
This 2\% is a proportion from the distribution. Only if we randomly
select from this population does 0.02 become a probability. Defining
percentiles through hypothetical random selections is unnecessarily
cumbersome.

By emphasizing the sampling distribution over hypothetical repeated
sampling, we can compare the distribution of test statistics with
familiar examples like poker hands. The distribution of 5-card poker
hands parallels the distribution of samples of size $n$ from a population
of $N$ individuals. While $N$ is typically unknown, both the number
of samples and number of poker hands are finite. Just as poker players
use hand rankings to assess winning chances, researchers use test
statistic values to assess the plausibility of the null hypothesis.

A key difference is that the 52-card deck is known while the population
is not. The null hypothesis specifies a model for the population.
Let $\mathcal{K}$ be the set of distinct population values. The population
distribution can be visualized as an area histogram over $\mathcal{K}$,
where each rectangle's area equals the proportion of that value in
the population.

The models are specified not by listing proportions for each value, but
by choosing from a family of distributions on a set $\mathcal{X}\supseteq\mathcal{K}$.
When $\mathcal{X}$ is countable, the model provides an area histogram;
when uncountable, a density curve. (See the appendix for details.)

The sampling distribution involves no randomization; it is determined
by the population distribution. Under the assumption that some distribution
in the family approximates the true population distribution, the corresponding
sampling distribution approximates the true sampling distribution.
For this approximation to describe the distribution of observed samples, the sample must
have been obtained by random sampling.

\section{Discussion}

\subsection{Practical Application}

For distributions like SAT scores or heights, standardized scores
and percentiles supplement the original values since the units are
meaningful. Test statistics lack units, so standardized scores can
replace them entirely. Software could report $\zeta_{obs}$ for two-tailed
tests instead of observed statistics from the $t$ and normal distributions
and $s_{obs}$ for one-tailed tests instead of the observed statistics
from the $F$ and chi-squared distributions. 

Common critical values derive easily from $s=3.32$ (the 10th percentile).
Critical values for $\alpha=0.05,0.025,0.01,$ and $0.005$ are $s=4.32,5.32,6.64,$
and $7.64$ respectively. Note that $6.64=3.32+3.32$, showing that
$\alpha=0.01$ is a factor of 10 smaller than $\alpha=0.10$ since
$(2^{-3.32})^{2}=2^{-6.64}$. Using critical values emphasizes the
sampling distribution rather than hypothetical repeated sampling.

Since p-values can be computed as $2^{-s_{obs}}$ for one-tailed tests
and $2^{-|\zeta_{obs}|}$ for two-tailed tests, reporting them is
optional. 

A key advantage of $s$-values over p-values appears when combining
evidence from independent studies. Fisher's method combines p-values
using the sum of their natural logarithms (\citealp[pp. 99-101,][]{Fisher2006-qm}).
Since $-\ln(p)=s\ln(2)$, the combined $s$-value depends only on
the sum of individual $s$-values.

Consider three randomized controlled trials examining Mediterranean
diet effects on cardiovascular health. Each trial compared the same
diet intervention to standard advice in adults with elevated cardiovascular
risk, measuring systolic blood pressure (BP) and low-density lipoprotein
(LDL) cholesterol.

The p-values for no BP reduction are 0.10, 0.09, and 0.08, while p-values
for no LDL reduction are 0.13, 0.07, and 0.06. Which outcome shows
stronger evidence against the null? The p-values alone don't make
this obvious.

Converting to $s$-values: BP reduction shows $s=3.3,3.5,$ and $3.6$
(sum=10.4), while LDL reduction shows $s=2.9,3.8,$ and $4.1$ (sum=10.8).
Since 10.8>10.4, these studies provide stronger evidence against the
null for LDL reduction.

\subsection{Theoretical Properties}

The semi-tail transformation belongs to the extensive group of monotone
functions, so its theoretical justification merits discussion.

Logarithms play a fundamental role in distributional theory. Entropy
of a discrete distribution on $\mathcal{X}$ equals the expected value
of $-\log m(x)$, where $m(x)$ is the measure assigned to $x$. Kullback-Leibler
divergence from $m_{1}$ to $m_{2}$ equals the expected value of
$\log m_{1}-\log m_{2}$. The log likelihood and its derivative, the
score function, are central to statistical inference.

While these definitions typically use natural logarithms, information
theory often uses base 2. Entropy in base 2 yields bits; with natural
logarithms, nats. Semi-tail units use base 2 logarithms but aren't
bits---the logarithm applies to distribution tails, not individual
points. Unlike entropy or Kullback-Leibler divergence, tail definitions
require an ordering on $\mathcal{X}$.

Tail areas are central to Bahadur efficiency and related measures
(\citealp{Bahadur1960} and Chapter 10 in \citealp{Serfling2001-gh}).
Let $T_{n}$ be the test statistic distribution for samples of size
$n$ from distribution $m$ (differing from null distribution $m_{\circ}$).
Define $B_{n}=b_{n}(T_{n})$ where: 
\begin{eqnarray}
b_{n}(t) & = & -2\log P_{m}(T_{n}\ge t)\nonumber \\
 & = & (2\log2)s_{n}(t).
\end{eqnarray}

If the tail area decays exponentially fast with probability 1, the
Bahadur exact slope is: 
\begin{eqnarray}
\beta & = & \lim_{n\rightarrow\infty}n^{-1}B_{n}\nonumber \\
 & = & (2\log2)\lim_{n\rightarrow\infty}n^{-1}S_{n}
\end{eqnarray}

The decay rate in semi-tail units, $\sigma=\lim_{n\rightarrow\infty}n^{-1}S_{n}$,
relates to the Bahadur slope by $\beta=(2\log2)\sigma$. Since Kullback-Leibler
divergence bounds $\beta$: 
\begin{eqnarray*}
\beta & \le & 2K(m,m_{\circ})\\
\sigma & \le & (\log2)^{-1}K(m,m_{\circ})=K_{2}(m,m_{\circ})
\end{eqnarray*}
where $K$ uses nats and $K_{2}$ uses bits.

The Bahadur efficiency of test sequence $\{T_{n}^{(1)}\}$ relative
to $\{T_{n}^{(2)}\}$ at $m$ equals: 
\[
\epsilon_{12}(m)=\frac{\beta_{1}}{\beta_{2}}=\frac{\sigma_{1}}{\sigma_{2}}.
\]

This efficiency is interpreted as the ratio of sample sizes needed
to achieve the same power: if $\epsilon_{12}(m)=2$, then test 2 requires
twice the sample size of test 1 for equal power.

For equal sample sizes, if $\sigma_{1}-\sigma_{2}>0$, then $\{T_{n}^{(1)}\}$
is more efficient. The difference $\sigma_{1}-\sigma_{2}$ is the
\emph{semi-tail efficiency}, and $2^{-(\sigma_{1}-\sigma_{2})}$ is
the \emph{tail area efficiency}. For example, if $\sigma_{1}-\sigma_{2}=0.15$,
the tail area efficiency is $2^{-0.15}=0.90$. For large $n$, the
more efficient test moves samples 10\% farther into the tail (0.15
semi-tail units), effectively reducing p-values by 10\%.

\subsection{Terminology}

While some statistical terms like ``mean'' and ``median'' align
with everyday usage, most technical terms should be treated as specialized
vocabulary. The semi-tail $s$-value has been called ``binary surprisal''
\citep{Rafi2020-qe}. However, surprise is subjective in everyday
use, while the $s$-value objectively measures an observation's position
in the distribution. Moreover, surprise implies belief in the null
hypothesis, which researchers typically question. Like ``probability,''
the term ``surprise'' conflates epistemic uncertainty about the
population with data's stochastic properties.

Terminology should describe theoretical definitions and computational
procedures rather than specific applications. We use ``distribution
variable'' rather than ``random variable'' for this reason. While
introductory texts (e.g., \citet{Hogg1978-ev}, or \citet{Casella2001})
emphasize hypothetical randomization, the measure-theoretic definition
requires none. The distribution of poker hands can be computed exactly
through combinatorics---repeated sampling would only approximate it.
Similarly, test statistic distributions are defined measure-theoretically
but could be approximated through repeated sampling. For both poker
hands and test statistics, a single randomization justifies using
the distribution to interpret observations. See the appendix and \citet{Vos2025-ai}
for details.

\section{Conclusion}

Semi-tail units provide a universal scale for hypothesis testing that
addresses longstanding challenges in statistical practice. By standardizing
all test statistics to a common scale where each unit represents a
halving of tail area, researchers can directly compare results across
different tests, easily combine evidence from multiple studies, and
interpret extremeness without consulting distribution-specific tables.
The connection to Bahadur efficiency reveals that semi-tail units
are not merely a convenient transformation but the natural scale for
measuring statistical efficiency. As statistical software increasingly
moves beyond p-values, implementing $s$- and $\zeta$-values offers
a principled path forward that preserves the useful aspects of hypothesis
testing while providing more intuitive interpretation.

\bibliography{vos_mzeta2}

\section*{Appendix}

The purpose of this appendix is to encourage the reader to think of
the symbol $X$ without invoking the idea of randomization even though
standard terminology denotes this a random variable. When there is
a family of distributions for $X$, members in the family are distinguished
using real numbers called parameters. For example, $\theta_{1}\not=\theta_{2}$
identify two distributions in the family. We identify the distributions
directly using the notation $m_{1}$ and $m_{2}$. 

Let $\mathcal{X}$ be a measurable space and $m_{\!\mathcal{A}}$
a measure on $\mathcal{A}=\sigma(\mathcal{X})$ such that $m_{\mathcal{A}}(\mathcal{X})=1$.
When $m_{\!\mathcal{A}}$ is dominated by measure $\mu$, there exists
a density function $m$ on $\mathcal{X}$ such that for all $A\in\mathcal{A}$
\[
m_{\!\mathcal{A}}(A)=\int_{A}m(x)\mu(dx).
\]

\begin{defn}
$X\subset\mathcal{X}\times(0,\infty)$ defined by 
\[
X=\{(x,m(x)):x\in\mathcal{X}\}
\]
is a \emph{distribution variable on} $\mathcal{X}$. $X$ is \emph{discrete}
if $\mathcal{X}$ is countable and \emph{simple} if $\mathcal{X}$
is finite with $Nm(x)$ an integer for all $x\in\mathcal{X}$ and
some integer $N$. 
\end{defn}
Simple distributions are fundamental to statistical inference. All
real populations are finite, making their distributions simple. Other
distributions serve as convenient approximations.

Real-valued simple distributions can be visualized using area histograms.
For each $x\in\mathcal{X}$, construct a rectangle with area $m(x)$
centered at $x$. Adjacent rectangles meet at midpoints between values.
More general distributions extend this by allowing non-rational areas
and additional rectangles, potentially creating smooth curves (densities).

Simple distributions also model physical randomization. Since all
physical populations are finite, sampling from infinite sets lacks
clear meaning.

In simple distributions, each element pairs with a proportion. A standard
52-card deck forms a simple distribution with $N=52$. The distribution
of 5-card poker hands has $N=\genfrac(){0 pt}{2}{52}{5}$. While the card deck
corresponds to a physical object, the poker hand distribution does
not. Nevertheless, the proportion $m(r)$ for rank $r$ equals the
probability of drawing that rank from a well-shuffled deck.

Statistical inference parallels this structure. Consider a population
with distinct values $\mathcal{X}_{pop}$ and let $X_{pop}$ be the
simple distribution of relative frequencies. While $X_{pop}$ corresponds
to the physical population, the sampling distribution $Y_{pop}$ for
samples of size $n$ does not. When sample $y\in\mathcal{Y}_{pop}$
is obtained by randomization, its probability equals the proportion
assigned by $Y_{pop}$.

Unlike poker, statistical inference faces unknown $X_{pop}$ and $Y_{pop}$.
We address this by specifying a family of distributions, assuming
some member approximates $X_{pop}$ with corresponding sampling distribution
approximating $Y_{pop}$.

Though populations have exact simple distributions, working with more
general distributions proves easier. Rather than specifying unknown
$N_{pop}$, we consider limits as $N\to\infty$, allowing non-rational
values for $m(x)$. When $\mathcal{X}_{pop}$ contains measurements
(necessarily rational), we may approximate using distributions on
real intervals for theoretical simplicity. 


\end{document}